\documentstyle[epsf,11pt]{article}
\newcommand{\bea}{\begin{eqnarray}}
\newcommand{\eea}{\end{eqnarray}}

\def\lsim{\mathrel{\vcenter{\hbox{$<$}\nointerlineskip\hbox{$\sim$}}}}

\begin{document}
\begin{titlepage}
\thispagestyle{empty}
%23.1.2001
%\vspace{0.2cm}
\vspace{0.2cm}

\title{{\normalsize \hfill HIP-2000-42/TH}\\[20pt]
 Single superparticle production via $\gamma\gamma$ collision
        with explicit R-parity violation
\footnote{The authors thank the Academy of Finland
(project numbers 163394 and 48787) for financial support.}}
\author{M. Chaichian$^{a,b}$, K. Huitu$^a$ and Z.-H. Yu$^{a,b}$\\
$^a$Helsinki Institute of Physics\\
$^b$Department of Physics, University of Helsinki\\
P.O.Box 9, FIN-00014 Helsinki, Finland}

\date{}

\maketitle

%\vskip 12mm
\vspace*{2truecm}

\begin{center}\begin{minipage}{5in}

\begin{center} ABSTRACT\end{center}
\baselineskip 0.3in

{\small We study the single production of scalar neutrinos or
charginos via $\gamma\gamma$ collision in an R-parity ($R_{p}$) violating
supersymmetric model.
It may be possible to detect a 
sneutrino or a chargino at a Linear Collider (LC)
in $\gamma \gamma$ operation mode, as a test of
supersymmetry and $R_p$-violation.
Because of the clean background in LC,
stringent constraints on $R_p$ violating
parameters can be obtained even if the process
cannot be observed at the future Linear Collider.} \\

\vskip 10mm

%{~~~~PACS number(s): 13.65.+i, 13.88.+e, 14.65.-q, 14.80.Dq, 14.80.Gt}
\end{minipage}
\end{center}
\end{titlepage}

\baselineskip=0.36in

\eject
\rm
\baselineskip=0.36in
\baselineskip=0.18in
\begin{flushleft} {\bf I. Introduction} \end{flushleft}
\par
\noindent
One of the main aims of a Linear Collider (LC) is detecting
supersymmetry \cite{s1}. Because of its clean background
compared with hadron colliders, LC can efficiently probe
new physics beyond the Standard Model (SM).
In addition to the $e^{+} e^{-}$ collider mode, the LC can, with the
advent of new collider techniques, produce
highly coherent laser beams being back-scattered with high luminosity
and efficiency at the $e^{+}e^{-}$ colliders \cite{s2}.
In this paper we will concentrate on $\gamma\gamma$ collisions.

The R-parity ($R_{p}=(-1)^{3B+L+2S}$, where $B$, $L$ and $S$ denote
the baryon number, lepton number and spin), which
is introduced to guarantee the $B$-
and $L$-conservation automatically, is conserved
in the minimal supersymmetric extension of the SM \cite{s3}.
With the discrete symmetry the lightest supersymmetric
particle (LSP) is stable and superparticles can only be pair
produced.
However, 
$R_p$ is not necessary in order to forbid fast proton
decay \cite{s4}.
Since in the $R_{p}$-violating ($\rlap/\! R_{p}$) models superparticles
can be singly produced and neutrinos get masses and mix \cite{s5},
it is a significant source of new physics.
Especially after the first signals for neutrino oscillations from
atmospheric neutrinos were
observed in Super-Kamiokande \cite{s6}, $R_{p}$-violation became
a good candidate to explain those experimental results.

The $R_p$-violation will introduce many processes forbidden
in the Standard Model. Thus, it is limited by the low energy
experiments \cite{s8}\cite{s9}.
Since the single production of superparticles is
admitted in the $R_p$-violating model, it can lower
superparticle production threshold.
If Nature favors the $R_p$-violation,
then the single production of superparticles may be the first
sign of supersymmetry.
In this work we will consider the single production of
scalar neutrinos and the lightest chargino in $\gamma\gamma$ collisions.

Detection of $R_p$-violation at the lepton colliders has
been considered both indirectly \cite{a1} and directly by
detecting the $R_p$-violating decay of superparticles produced at
the lepton colliders \cite{a2}, and
by producing superparticles singly \cite{s10}.
The single production of scalar neutrinos from $e^+e^-$
collision has  been considered in \cite{s10},
where the $L$-violating parameters $\lambda$ involving
light flavors will dominate
the process. 
The processes $e^{+}e^{-}\rightarrow
\tilde{\chi}^0 \nu, \, \tilde{\chi}^{\pm} l^{\mp}$ were also considered
in \cite{s10}.
However, $\gamma\gamma$ collision could introduce
the $R_p$-violating parameters
involving heavy flavors. 
These parameters could be much larger than
those involving light flavors with an assumption of
family symmetry \cite{s11}, and also introduce
other $L$-violating parameters $\lambda^{'}$ in the processes.
A $\gamma\gamma$ resonance can be probed over a wide
mass region, even before production in direct $e^{+}e^-$ collision,
which is only sensitive at the center of mass energy (c.m. energy)
of colliders.  The resonant sneutrino production at Large Hadron
Collider (LHC) has already been considered in Ref. \cite{s12}.

In section 2, the cross section of the process 
$\gamma\gamma (\rightarrow \tilde{\nu}) \rightarrow
\tilde{\chi}^{\pm}+\l^{\mp}$
is calculated.
In section 3 the signals of the processes
$$
\gamma\gamma\rightarrow \tilde\nu , \eqno{(1.1)}
$$$$
\gamma\gamma  \rightarrow
\tilde{\chi}^{\pm}+\l^{\mp}\eqno{(1.2)}
$$
are considered.
Our conclusions are given in section 4 and some details of
the expressions are listed in the appendix.

\begin{flushleft} {\bf 2. Production of $\tilde\nu$ and
$\tilde\chi^\pm$
with explicit R-parity violation }\end{flushleft}
\par
\noindent
All renormalizable supersymmetric $\rlap/\! R_{p}$ interactions can
be introduced in the superpotential \cite{s9}:
$$
\begin{array} {lll}
    W_{\rlap/\! R_{p}} & =\frac{1}{2}
\lambda_{[ij]k} L_{i}.L_{j}\bar{E}_{k}+\lambda^{'}_{ijk}
L_{i}.Q_{j}{\bar D_{k}}+\frac{1}{2}\lambda^{''}_{i[jk]}
\bar{U}_{i}\bar{D}_{j}\bar{D}_{k}+\epsilon _{i} L_{i} H_{u}.
\end{array}
\eqno {(2.1)}
$$
where $L_i$, $Q_i$ and $H_u$ are SU(2) doublets containing lepton, quark
and Higgs superfields respectively, $\bar{E}_j$ ($\bar{D}_j$, $\bar{U}_j$)
are singlet lepton (down-quark and up-quark) superfields,
and $i,j,k$ are generation indices.
Square brackets on them denote
antisymmetry in the bracketted indices.

We will consider only trilinear terms in this paper.
In order to avoid fast proton decay, it is necessary that \cite{s9}
$$
|(\lambda~{\rm or}~\lambda^{'}) \lambda^{''}|<
10^{-10}\left(\frac{\tilde{m}}{100\;{\rm GeV}}\right)^{2},
\eqno {(2.2)}
$$
where $\tilde{m}$ is the mass of a squark or a slepton.
We will consider only the $L$-violating terms in
our calculations. We also assume that the parameters $\lambda$
and $\lambda^{'}$ are real. 

One-loop corrections (the ones
corresponding to $\lambda$ terms are
shown in Fig.1,
and contributions
from $\lambda^{'}$ terms are similar)
to $\gamma\gamma \rightarrow \tilde{\chi}^{-} l^{+}$
can be split into the following components:
$$
M = \delta M_{s}+ \delta M_{v} + \delta M_{b},
\eqno {(2.3)}
$$
where $\delta M_{s}$, $\delta M_{v}$ and $\delta M_{b}$
are the one-loop amplitudes corresponding to the self-energy, vertex,
and box correction diagrams, respectively.
Since the proper
vertex counterterm should cancel with the counterterms of the external
legs in this case, we do not need to deal with the ultraviolet
divergence. 
We simply
sum over all (unrenormalized) reducible and irreducible diagrams and
the result is finite and gauge invariant.

\begin{figure}[t]
\leavevmode
\begin{center} 
\mbox{\epsfxsize=6.truecm\epsfysize=6.truecm\epsffile{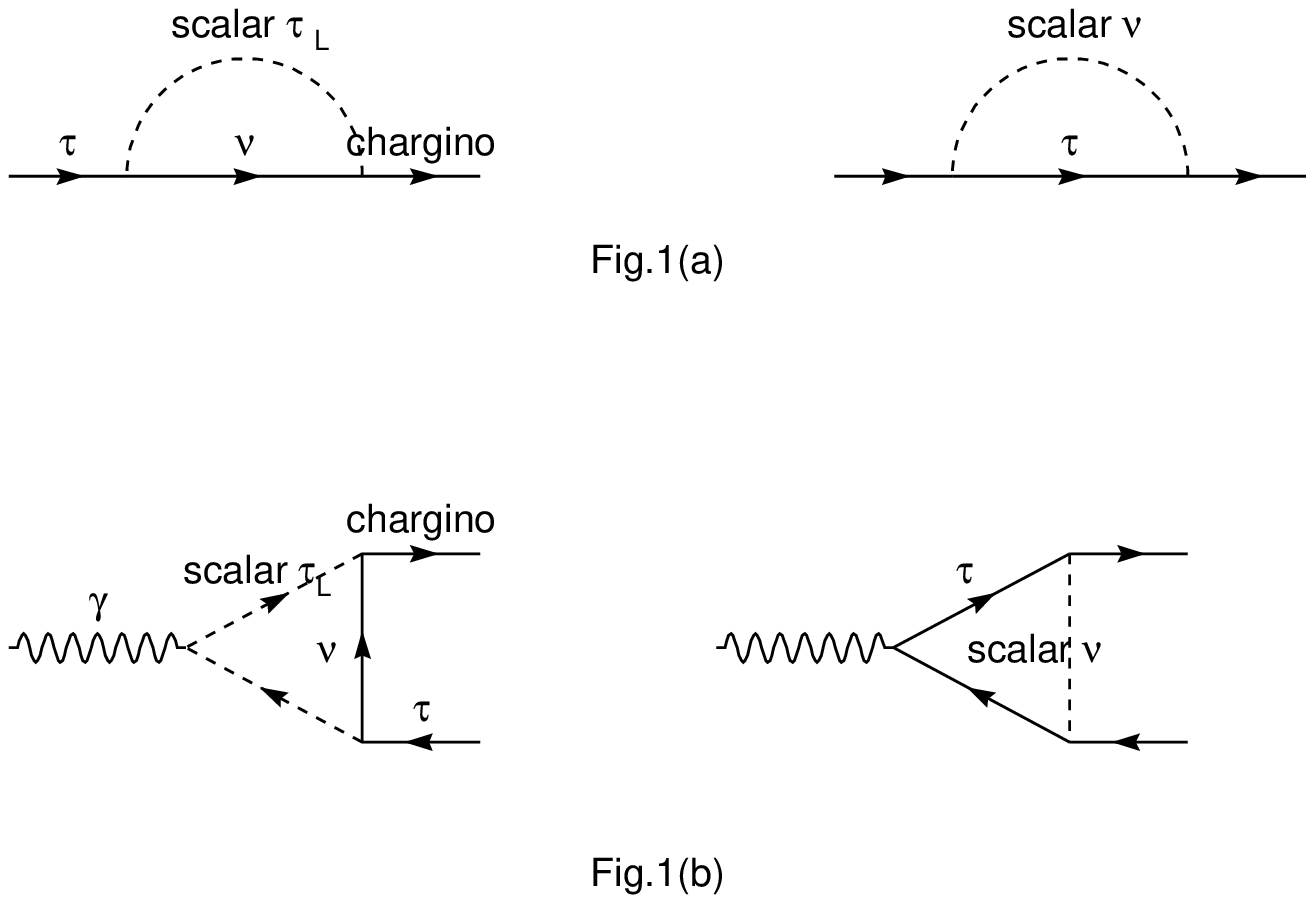}}
\mbox{\epsfxsize=6.truecm\epsfysize=6.truecm\epsffile{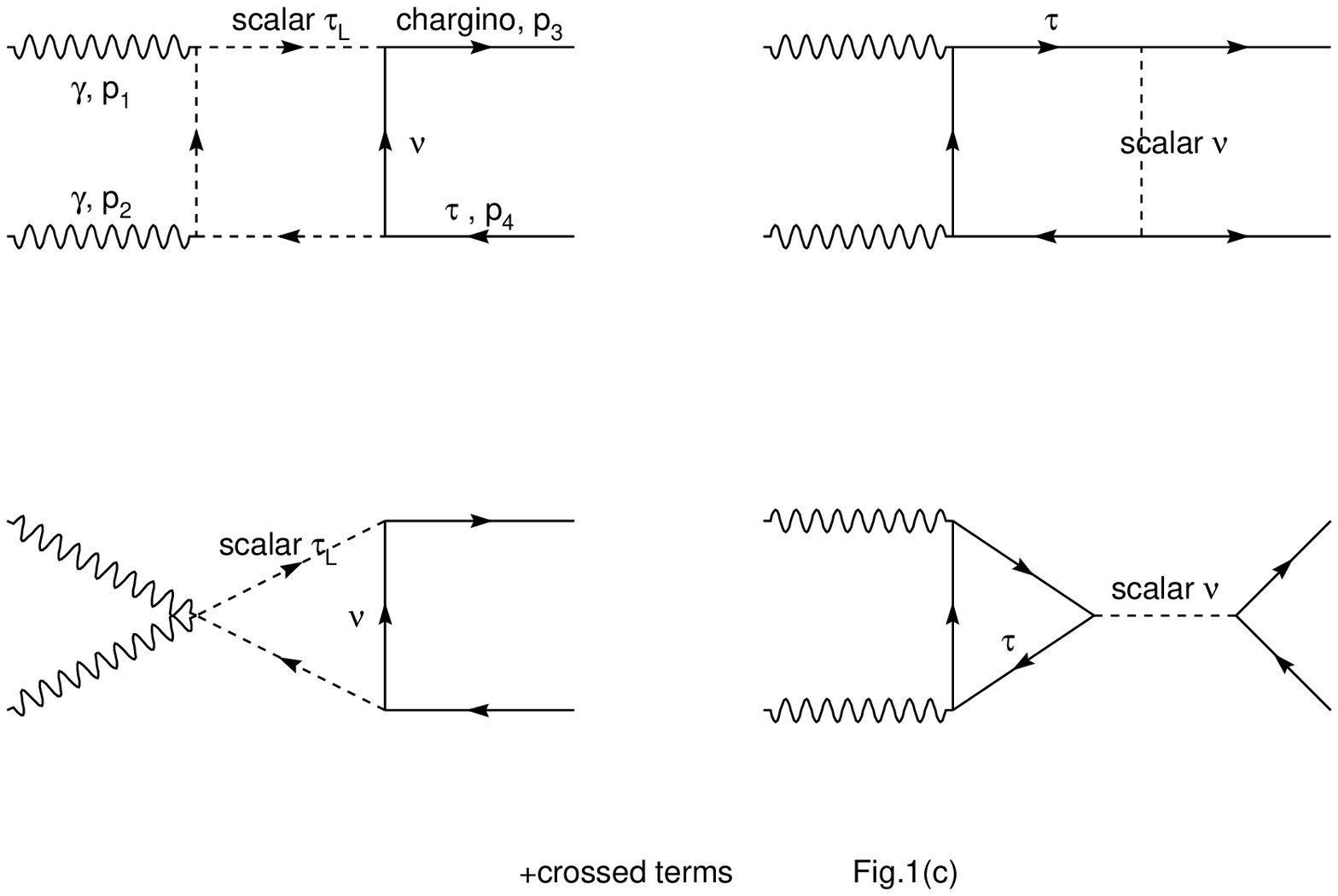}}
\end{center}
\caption{\label{fig1}
         Feynman diagrams of $\gamma\gamma\rightarrow \tilde{\chi}^{-}   
         \tau^{+}$.  Here
        (a) are self-energy diagrams,
        (b) vertex diagrams, and
        (c) box, quartic interaction and triangle diagrams.}
\end{figure}

Thus, we can get the amplitude of $ \gamma\gamma \rightarrow
\tilde{\chi}^{+}l^{-}$.
Collecting the terms together, we obtain the total cross section for the
subprocess $\gamma\gamma \rightarrow \tilde{\chi}^{\pm} l^{\mp}$:
$$
\hat{\sigma}(\hat{s}) = \frac{1}{16 \pi \hat{s}^2 }
             \int_{\hat{t}^{-}}^{\hat{t}^{+}} d\hat{t}\;
{\overline{\sum}_{spins}^{}}
             |M|^{2},
\eqno{(2.4)}
$$
where $\hat{t}^{\pm}=\frac{1}{2}\left[
(m_{\tilde{\chi}^{\pm}}^2+m_l^2-\hat{s})\pm
\sqrt{\hat{s}
^2+m_{\tilde{\chi}^{\pm}}^4+m_l^4-2\hat{s} m_{\tilde{\chi}^{\pm}}^2-2\hat{s} m_l^2-2
m_{\tilde{\chi}^{\pm}}^{2} m_l^{2}} \right]$,
and the bar over summation means averaging over
initial spins. In order to obtain the observable cross
section
of single chargino production via $\gamma \gamma$ fusion
in an $e^{+}e^{-}$ collider, we need to fold the cross section of
$\gamma\gamma
\rightarrow \tilde\chi^{\pm}l^{\mp}$ with the photon luminosity,
$$
\sigma(s) = \int_{(m_{\tilde{\chi}^{\pm}}+m_l)/\sqrt{s}}^{x_{max}} dz
\frac{dL_{\gamma\gamma}}{dz}
              \hat{\sigma}(\hat{s}),
\eqno {(2.5)}
$$
where $\hat{s}=z^2 s$, $\sqrt{s}$ and $\sqrt{\hat{s}}$ are the $e^{+}e^{-}$
and $\gamma\gamma$ c.m. energies respectively.
$\frac{dL_{\gamma\gamma}}{dz}$
is the photon luminosity, which is defined as \cite{s2}
$$
\frac{dL_{\gamma\gamma}}{dz} = 2z \int_{z^{2}/x_{max}}^{x_{max}} \frac{dx}{x}
                                F_{\gamma /e}(x)F_{\gamma /e}(z^{2}/x),
\eqno {(2.6)}
$$
where the energy spectrum of the back-scattered photon is given by \cite{s2}
$$
F_{\gamma /e}(x) = \frac{1}{D(\xi)} [ 1-x+\frac{1}{1-x}-\frac{4x}{\xi
(1-x)}+
                   \frac{4x^{2}}{\xi^{2} (1-x)^{2}} ],
\eqno {(2.7)}
$$
and
\cite{s13}
$\xi=4.8$,
$x_{max}=0.83$ and $D(\xi)=1.8$.
\begin{flushleft} {\bf 3. Numerical results} \end{flushleft}

\noindent 
Let us first consider the production of sneutrinos. 
The cross section will only depend on the $\rlap/\! R_{p}$ couplings 
and masses of sneutrinos, but not on the other sparticle masses.
With an assumption of family symmetry discussed in \cite{s11}, 
$R_{p}$-violating couplings involving heavy flavors will be larger
than those involving light flavors.
Thus,
we will consider mainly couplings $\lambda_{23i}$, 
$\lambda'_{22i}$, $\lambda'_{23i}$, and $\lambda'_{3ij}$.
These are also experimentally least bounded.
The productions of $\tilde \nu_2$ and $\tilde \nu_3$ differ
qualitatively.
In the case of $\tilde\nu_2$, the production via
both $\lambda$ and $\lambda'$ terms are important, since either 
$\tau$-lepton or $b$-quark can circulate in the loop.
In the case of $\tilde \nu_3$ only $\lambda'$ term with $b$-quark in
the loop is significant. 

We plot in Fig. 2 cross sections for the single production of 
$\tilde \nu_2$ and $\tilde \nu_3$, using the largest allowed values
\cite{s9} of the relevant couplings, as well as using the close to 
smallest values observable for the couplings.
 In Fig. 2 (a), we show the cross section of $e^{+}e^{-}\rightarrow
\gamma\gamma \rightarrow \tilde{\nu}_2$ as a function of the mass of the
sneutrino $\tilde{\nu}_2$ for c.m. energy 500 GeV. 
The solid line corresponds to $\lambda_{233}=0.1$  and the dashed line to
$\lambda_{233}=0.01$, respectively. 
It can be seen that the cross section is still 0.005~fb for 
$m_{\tilde{\nu}}=400$ GeV with $\lambda_{233}=0.1$, which is almost 
the present upper limit for $\lambda_{233}$ \cite{s9}. 
Even with a much smaller $R_p$-violating coupling,
$\lambda_{233}=0.01$, 
the sneutrino production cross section for sneutrinos lighter than  
95 GeV\footnote{The present lower limits for sneutrino
masses are $m_{\tilde\nu_2}>84$ GeV and $m_{\tilde\nu_3}>86$ GeV
\cite{lepc}.}
remains above 0.002 fb (corresponding to one event per year with
the luminosity 500 fb$^{-1}$ at c.m. energy 500 GeV).
Similarly, in Fig. 2 (b), we plot the
cross section of $e^{+}e^{-}\rightarrow \gamma\gamma \rightarrow
\tilde{\nu}_2$ 
with the coupling $\lambda^{'}_{233}=0.15$ (present upper
limit), corresponding to the solid line, and $\lambda^{'}_{233}=0.03$ 
corresponding to the dashed line.
If $m_{\tilde\nu}=90$ GeV and $\lambda^{'}_{233}=0.15$, a few tens of
events are produced per year with 500 fb$^{-1}$ luminosity.
The cross section remains above 0.002 fb with
$\lambda^{'}_{233}=0.15$ for $m_{\tilde{\nu}}$ less than 400 GeV.
In Fig. 2 (c), we plot the
cross section of $e^{+}e^{-}\rightarrow \gamma\gamma \rightarrow
\tilde{\nu}_3$ as a function of the sneutrino $\tilde{\nu}_3$ mass
for c.m. energy $500$ GeV with
the coupling of $\lambda^{'}_{333}=0.45$ (present
limit).  Since  $\lambda^{'}_{333}$ can be
much larger than  $\lambda^{'}_{233}$, the cross
section of $\tilde{\nu}_3$ production could be larger within the present 
limits.  

\begin{figure}[t]
\leavevmode
\begin{center} 
\mbox{\epsfxsize=7.truecm\epsfysize=7.truecm\epsffile{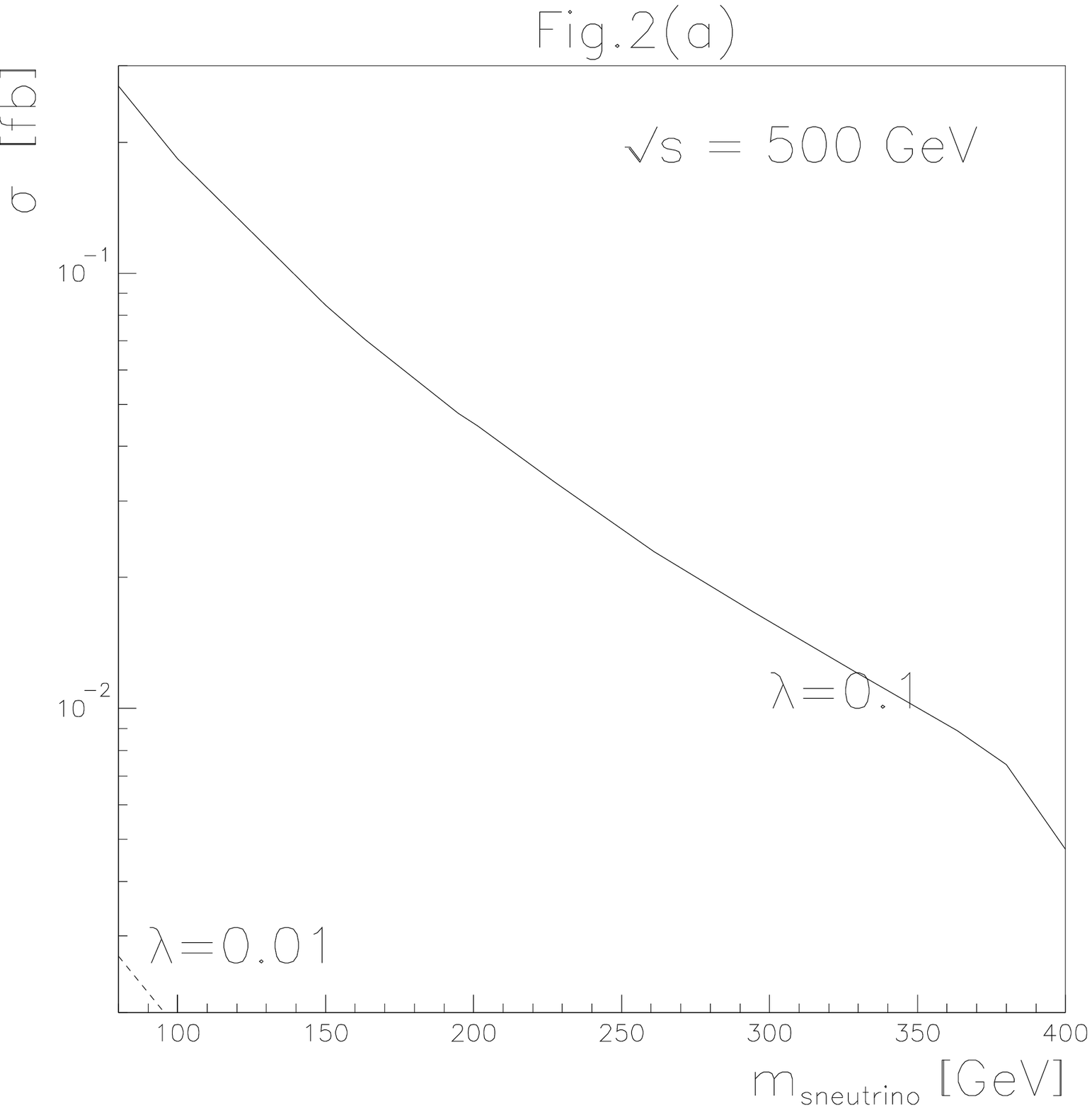}}
\mbox{\epsfxsize=7.truecm\epsfysize=7.truecm\epsffile{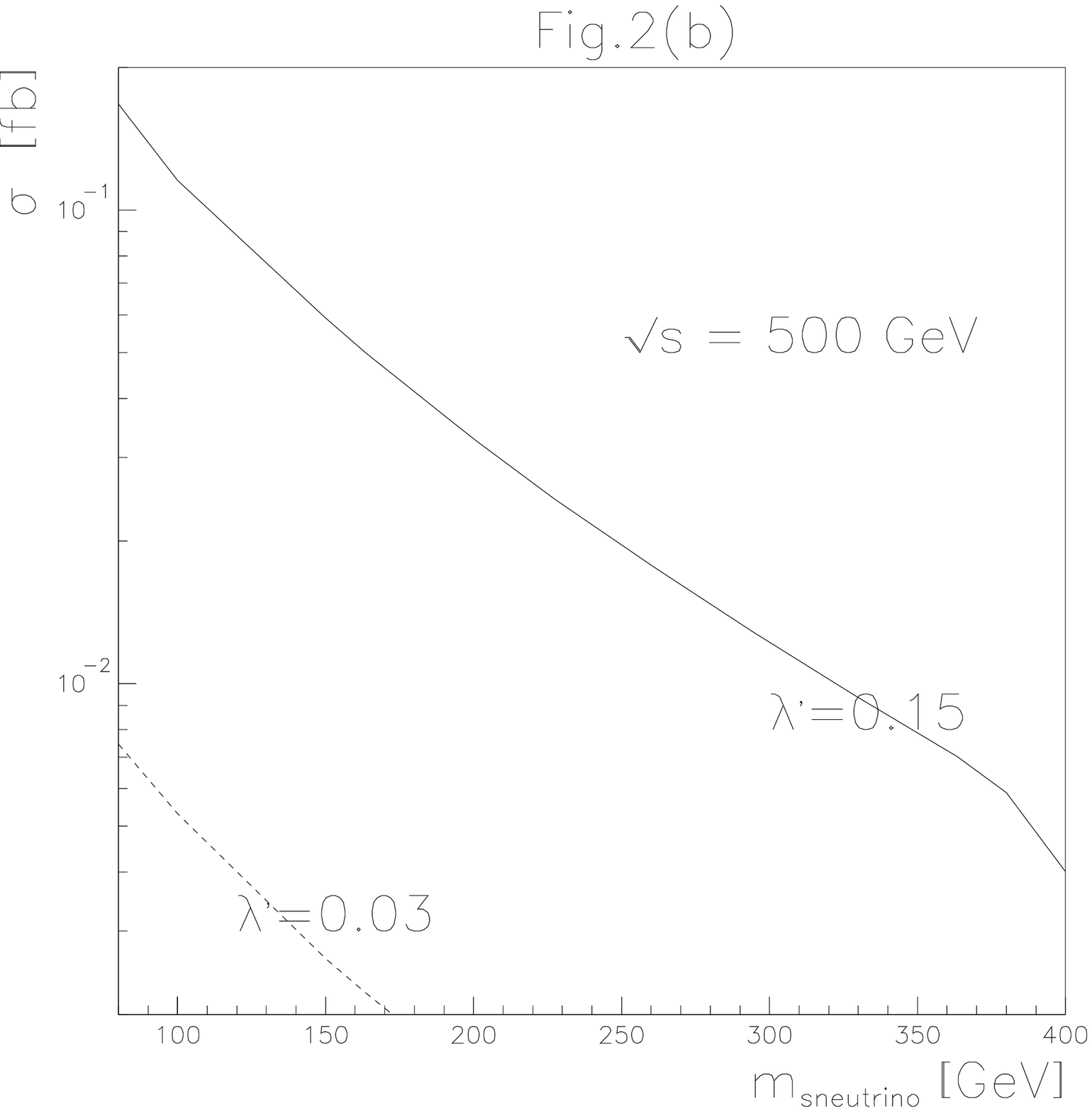}}
\mbox{\epsfxsize=7.truecm\epsfysize=7.truecm\epsffile{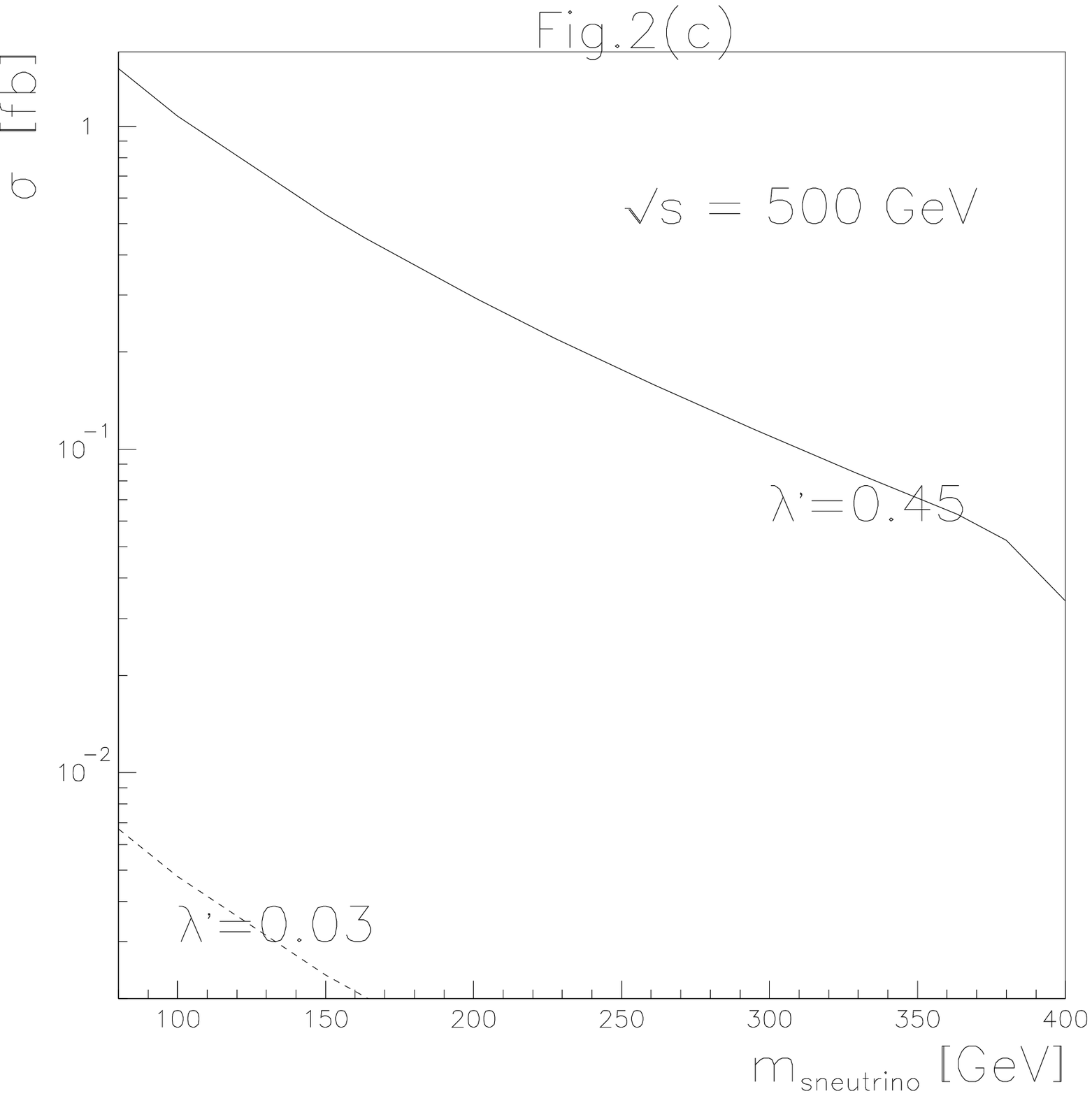}}
\end{center} 
\caption{\label{fig2}
As a function of the mass of the produced sneutrino, the cross section 
of
(a) $e^{+}e^{-}\rightarrow \gamma\gamma \rightarrow \tilde{\nu}_2$, 
where the solid line corresponds to $\lambda_{233}=0.1$,
          and the dashed line to $\lambda_{233}=0.01$,
(b) $e^{+}e^{-}\rightarrow \gamma\gamma\rightarrow \tilde{\nu}_2$,
where the solid line corresponds to $\lambda^{'}_{233}=0.15$ and
dashed line to $\lambda^{'}_{233}=0.03$, and
(c) $e^{+}e^{-}\rightarrow \gamma\gamma  \rightarrow \tilde{\nu}_3$,
where $\lambda^{'}_{333}=0.45$.
}  
\end{figure}

Next we consider the possible single production of charginos with $\rlap/\! 
R_p$ couplings. 
We assume here the minimal supergravity (mSUGRA) model, where
we take as our reference point $m_{0}=100$ GeV, $A_{0}=-100$ GeV, $\tan
\beta =3$ and $sign(\mu)=+$. The masses of sneutrinos and charginos
increase when we change $m_{1/2}$ from 100 GeV to 500 GeV, see Table 1.
Increasing $m_0$ would increase the masses of stau and sneutrino and
at the same time decrease the cross section to an unobservable level.

We plot the cross section of $\gamma\gamma \rightarrow
\tilde{\chi}_{1}^{\pm} l^{\mp}$ as a function of the chargino mass
$m_{\tilde{\chi}_1^{\pm}}$ for the c.m. energy 500 GeV in Figs. 3.
In Fig. 3 (a) we have $l=\mu$ and $\lambda_{233}=0.1$ and in
Fig. 3 (b) $l=\mu$ and $\lambda^{'}_{233}=0.15$.
In Fig. 3 (c) $l=\tau$ and $\lambda^{'}_{333}=0.45$.
It is seen in the cases (a) and (c)
that with the luminosity 500 fb$^{-1}$ at least one chargino
$\tilde{\chi}_1^{\pm}$ is produced if $m_{\tilde{\chi}_1^{\pm}} \lsim 200$ GeV.
Compared with the cross section
of pair production of charginos and neutralinos at the lepton colliders
\cite{s15}, the results are smaller if the charginos are light. However,
the single production of charginos can lower the threshold of production and
can provide us with a possible way to detect heavier charginos at lepton
colliders.
\begin{figure}[t]
\leavevmode 
\begin{center}
\mbox{\epsfxsize=7.truecm\epsfysize=7.truecm\epsffile{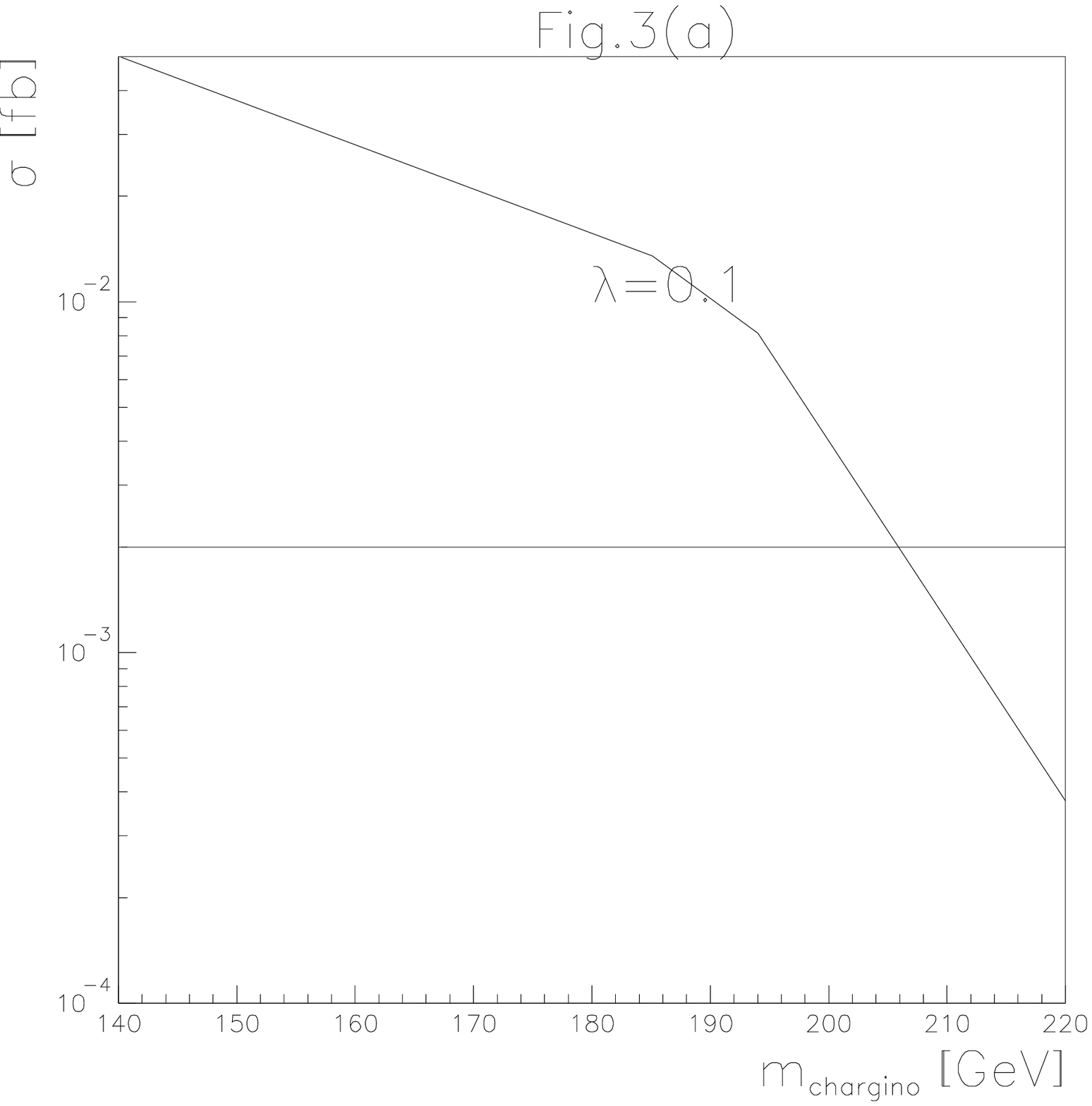}}
\mbox{\epsfxsize=7.truecm\epsfysize=7.truecm\epsffile{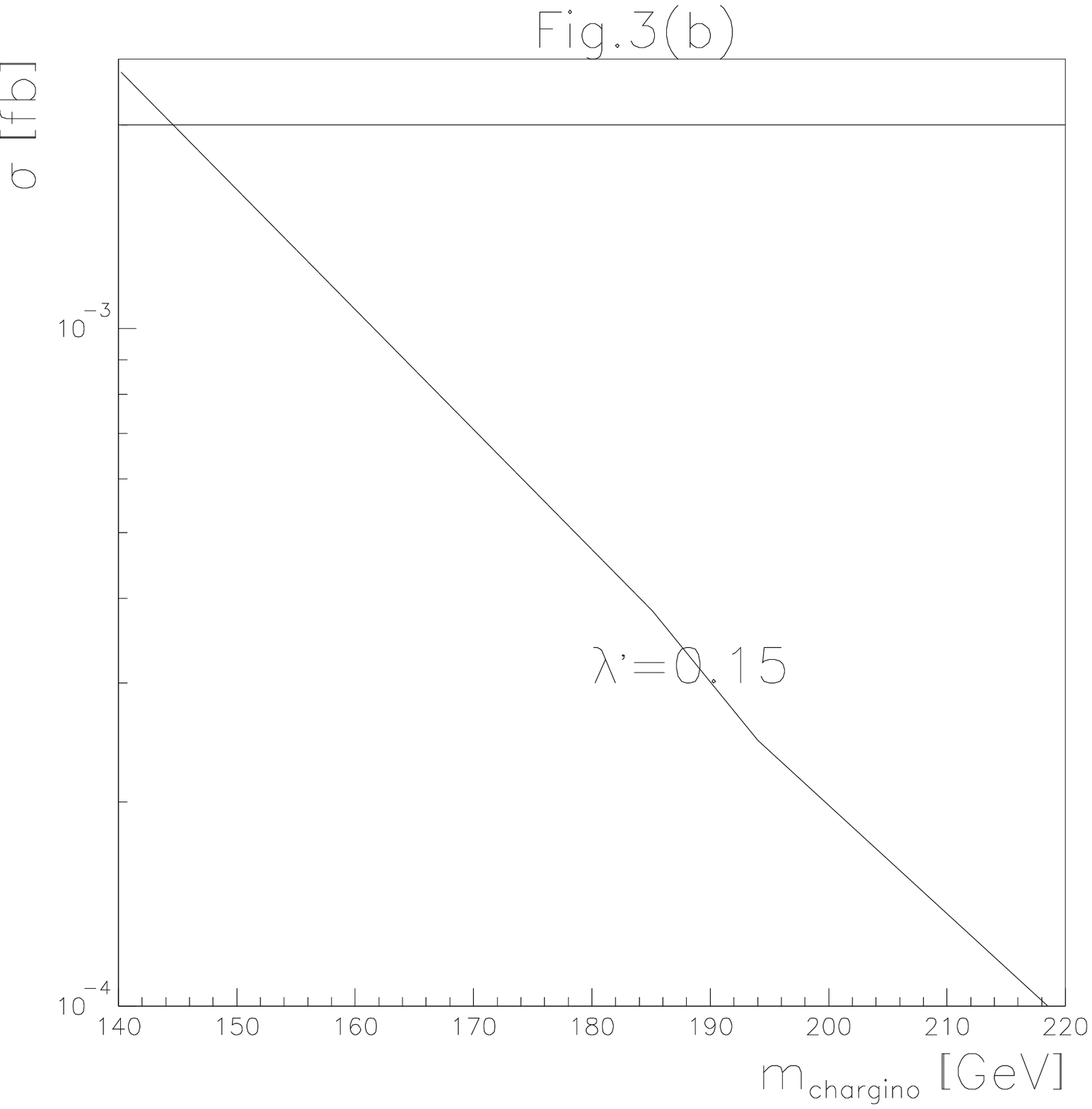}}
\mbox{\epsfxsize=7.truecm\epsfysize=7.truecm\epsffile{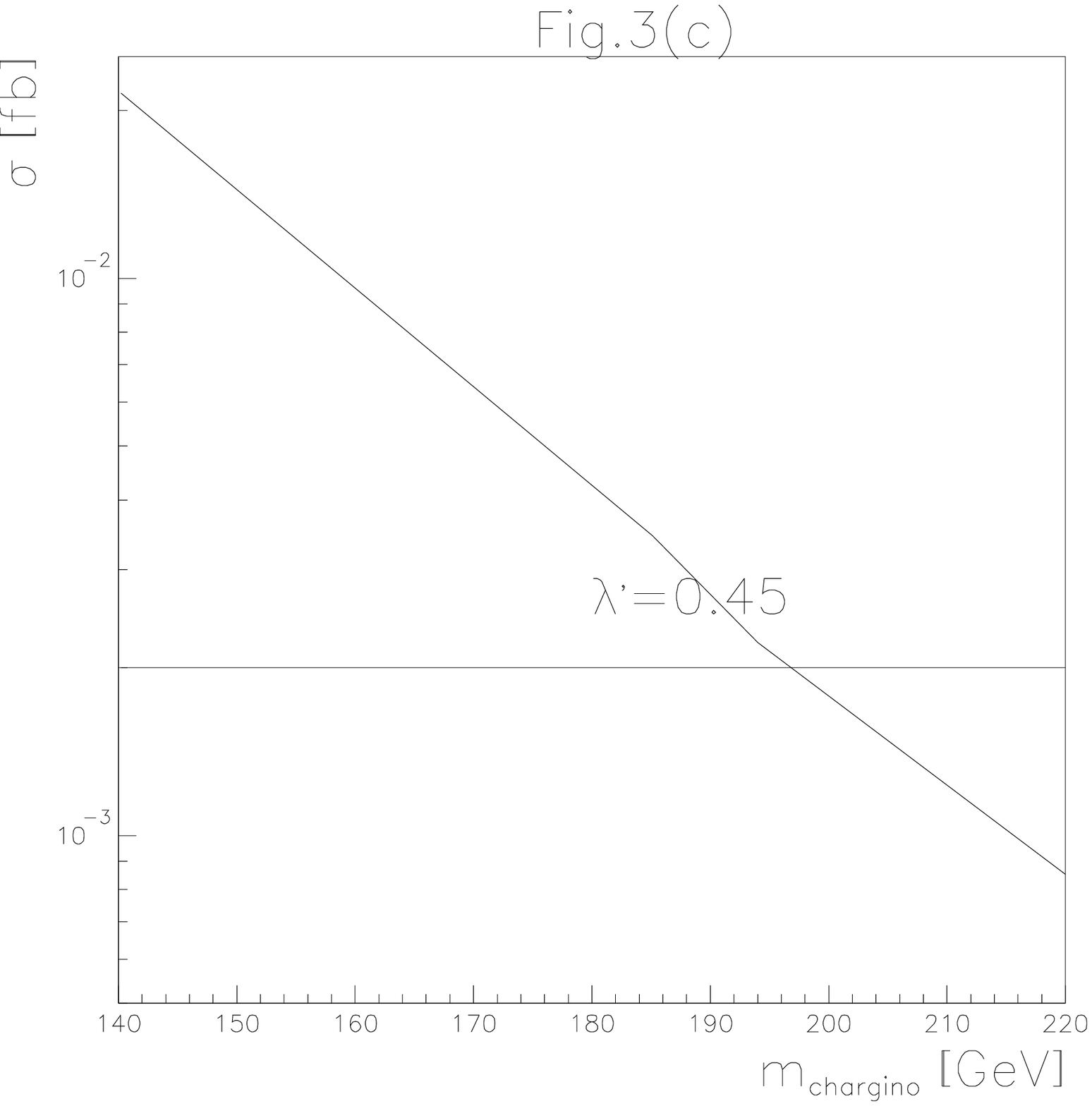}}
\end{center}
\caption{\label{fig3}
As a function of the mass of the chargino $\tilde\chi_1^\pm$,
the cross section of  $e^{+}e^{-}\rightarrow \gamma\gamma
        \rightarrow \tilde{\chi}^{\pm}_{1} l ^{\mp}$ at c.m.energy
        $E_{cm}=500$ GeV with
(a) $l=\mu$ and $\lambda_{233}=0.1$,
(b) $l=\mu$ and $\lambda^{'}_{233}=0.15$, and
(c) $l=\tau$ and $\lambda^{'}_{333}=0.45$.
The horizontal line corresponds to one event with 500 fb$^{-1}$.}

\end{figure}

In order to detect the signal events, the decays of sneutrinos and
charginos are of prime importantance. In the following we consider
the possible decay modes.

{\it Decay of sneutrinos.}
We consider here two essentially different possibilities for the
sneutrino decay: it may be the LSP, in which case the $\rlap /\! R_{p}$
decays dominate, or it may be that one or more of the neutralinos 
and charginos are lighter than the sneutrino.

If a sneutrino is the  LSP, it will decay through $R_p$-violating
terms. 
Assuming only one nonvanishing $R_p$ violating coupling, we can
conclude from the diagrams in Fig. 1 that there are no 
nondiagonal decays of sneutrinos if a sneutrino has been produced.
However, the decays with even small nondiagonal couplings may be 
experimentally important, since they induce flavor-changing
decay modes and thus provide spectacular final states.

With nonzero $\lambda_{233}$
coupling, the sneutrino $\tilde{\nu}_2$ will decay to $\tau^+\tau^-$, and with
nonvanishing $\lambda^{'}_{233}\;(\lambda^{'}_{333})$ coupling, the sneutrino 
$\tilde{\nu}_2 \;(\tilde{\nu}_3)$ will
decay to $b\bar{b}$.
If the nondiagonal $\lambda_{23i}=\lambda_{32i}$ are also
nonvanishing, the decays
$$
\tilde\nu_2\rightarrow\tau\mu,\,\tau e,\;\;
\tilde\nu_3\rightarrow\tau\mu,\,\mu e
$$
would be possible, or if the nondiagonal couplings
$\lambda^{'}_{23i}$ or $\lambda^{'}_{3ij}$ exist, the decay channels
$$
\tilde\nu_2\rightarrow b\bar s,\,\bar b s,\,b\bar d,\,\bar b d
$$
would be open.

If one of the lightest neutralinos $\tilde{\chi}^{0}_{1,2}$ or the 
chargino $\tilde{\chi}^{\pm}_{1}$ are lighter than sneutrinos,
then the $R_p$-conserving decay is also possible.
The possible decay channels are as follows:
$$
\begin{array}{lll}
\tilde{\nu}_i \rightarrow \tilde{\chi}_1^{\pm} l^{\mp}_i,\;
\tilde{\nu}_i \rightarrow \tilde{\chi}_{1,2}^{0} \nu _i.
\end{array}
\eqno(4.1)
$$
If kinematic space admits, sneutrino can also decay as follows 
$$
\begin{array}{lll}
\tilde{\nu}_i \rightarrow \tilde{l}^{\pm}_{iL} W^{\mp}.
\end{array}
\eqno(4.2)
$$

In our case, with sparticle masses shown as in Table 1,
only the decays in (4.1) are allowed.
In Figs. 4 and 5, we show the branching ratios of the
sneutrinos with parameters $\lambda_{23i}$ and
$\lambda^{'}_{33i}$ (with $\lambda{'}_{32i}$) dominating,
respectively. (The branching ratio of 
sneutrinos with parameters $\lambda^{'}_{233}$ and $\lambda{'}_{222}$
dominating will be the same as in Fig.5). 
We can see from the figures that the $R_p$-violating decay
of sneutrinos will be important for 
$\lambda_{23i}=\lambda_{2i3}=0.1$,
and even dominate when $\lambda_{333}^{'} (\lambda_{322}^{'})=0.45$.

If it turns out that the Higgs boson mass is close to the sneutrino 
mass, and $m_h\lsim 2m_W$, the sneutrino cannot be distinguished from 
the Higgs boson.
To see this explicitly, we have plotted in Fig. 6 the $\sigma\times Br
(\gamma\gamma\rightarrow h\rightarrow b\bar b,\,\tau\bar\tau,\, s\bar
s)$.
The cross section for the $b\bar b$ final state for a 100 GeV Higgs 
boson is more than 10 fb and for the $\tau\bar\tau $ final state is
around 0.7 fb, which are both larger than the sneutrino production
cross section.
If the $h$ and $\tilde\nu_i$ masses are clearly different, 
or if the Higgs mass is above $2m_W$ (in which case it decays dominantly
to a pair of gauge bosons) the situation changes,
and $\tau^+\tau^-$ or $b\bar b$ would provide good signals of 
$\tilde\nu_{2,3}$.
In the case of nondiagonal decay modes there is no background 
from the Higgs decay.
Thus, it is worth emphasizing that for similar mass sneutrinos and the Higgs
boson, the sneutrino decay modes
$ \tilde{\nu}\rightarrow \tau\mu,\,\tau e, 
\,bs $ are essential in order to detect sneutrinos.

\begin{figure}[t]
\leavevmode
\begin{center}
\mbox{\epsfxsize=7.truecm\epsfysize=7.truecm\epsffile{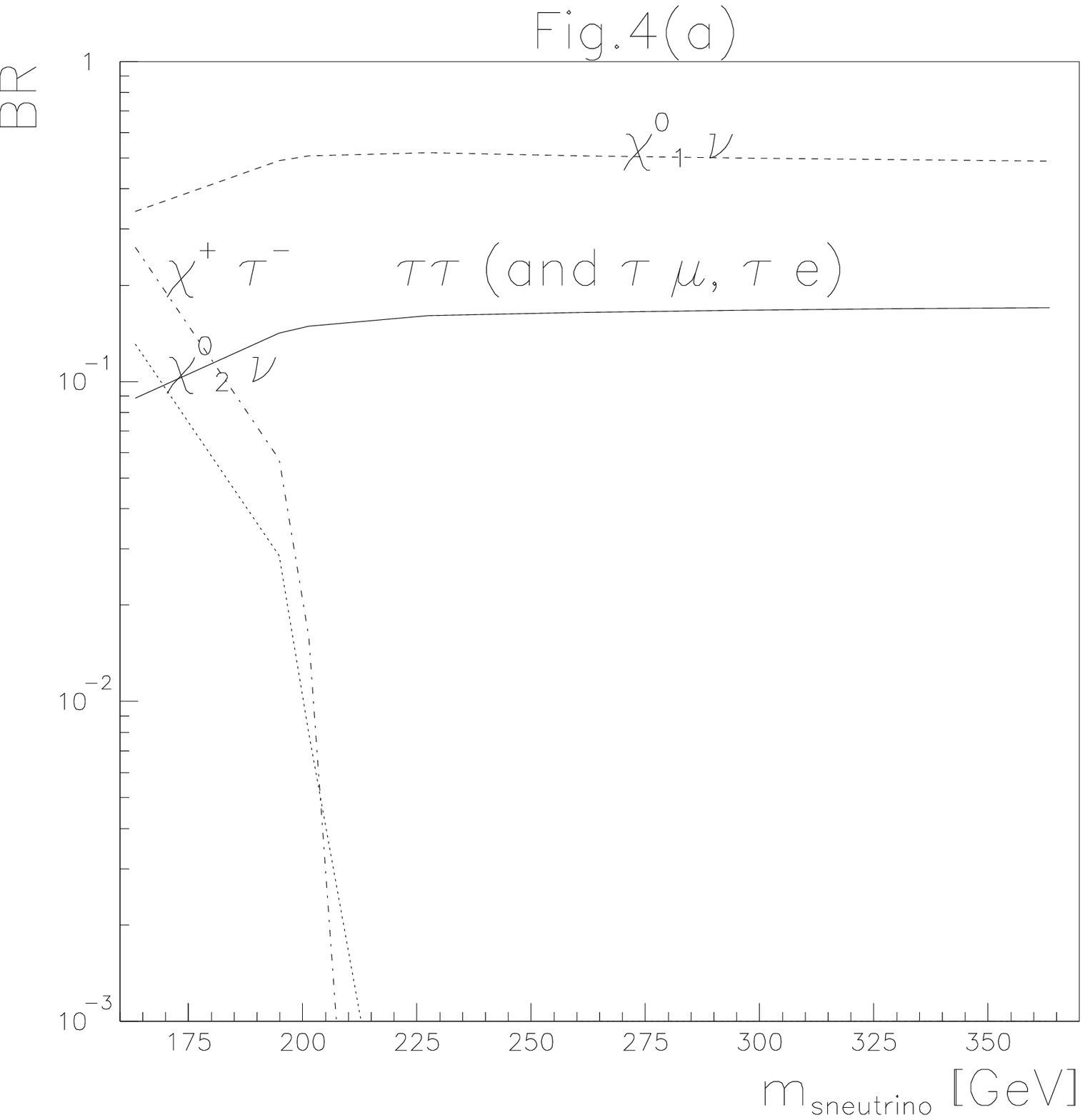}}
\mbox{\epsfxsize=7.truecm\epsfysize=7.truecm\epsffile{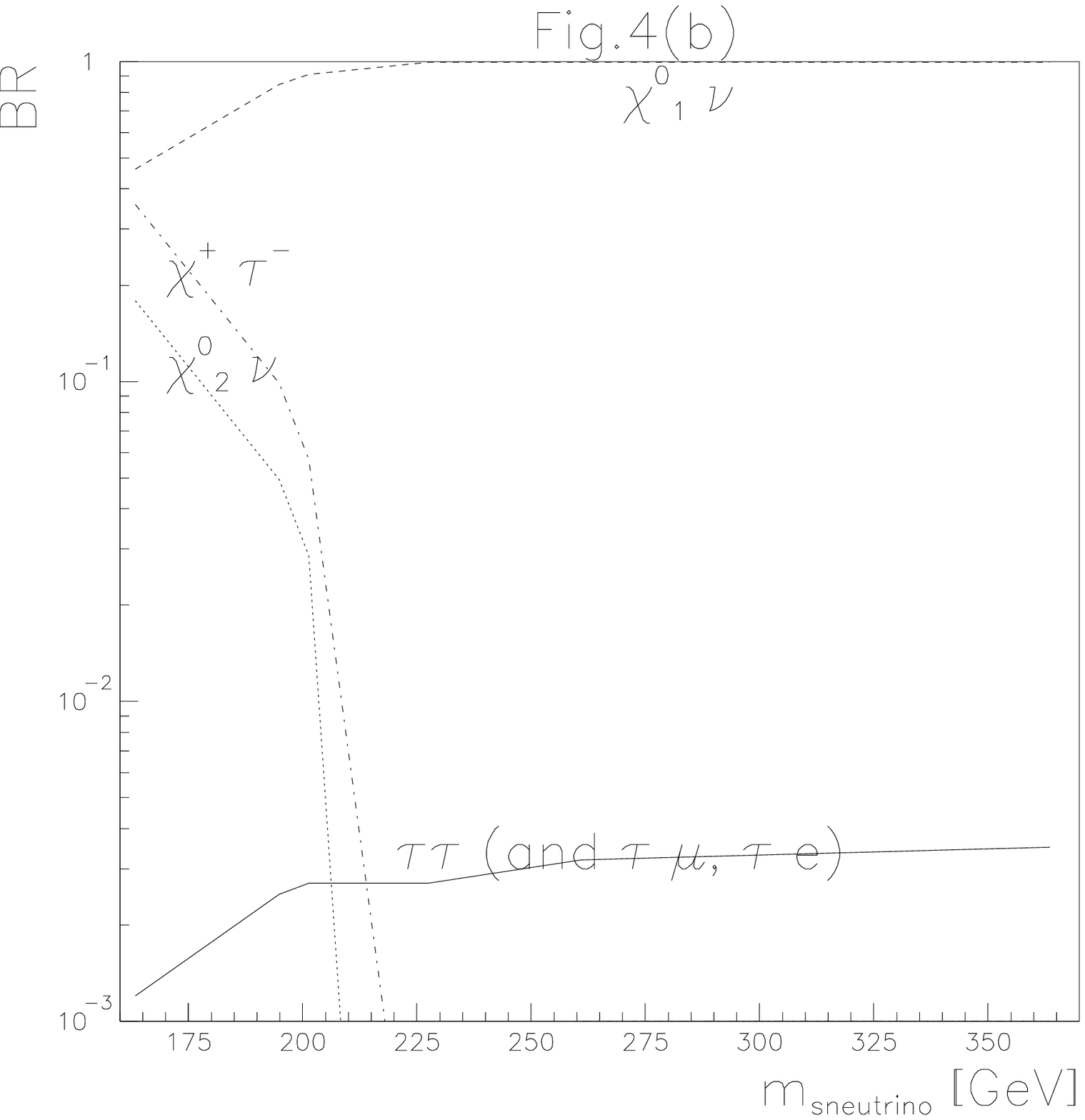}}  
\end{center}
\caption{\label{fig4}
Branching ratios of sneutrino $\tilde\nu_2$ with
        (a) $\lambda_{233}=0.1$ and with
        (b) $\lambda_{233}=0.01$.}
\end{figure}

\begin{figure}[t] 
\leavevmode
\begin{center}
\mbox{\epsfxsize=7.truecm\epsfysize=7.truecm\epsffile{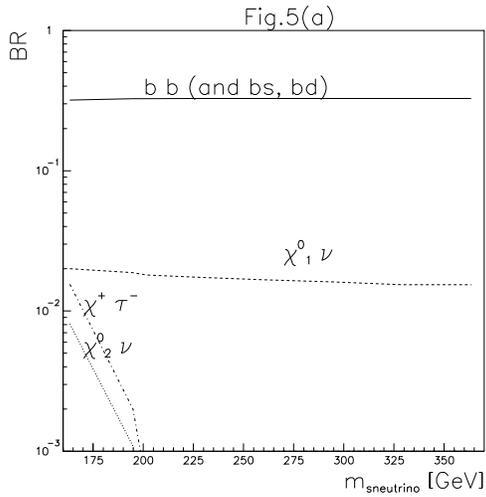}}
\mbox{\epsfxsize=7.truecm\epsfysize=7.truecm\epsffile{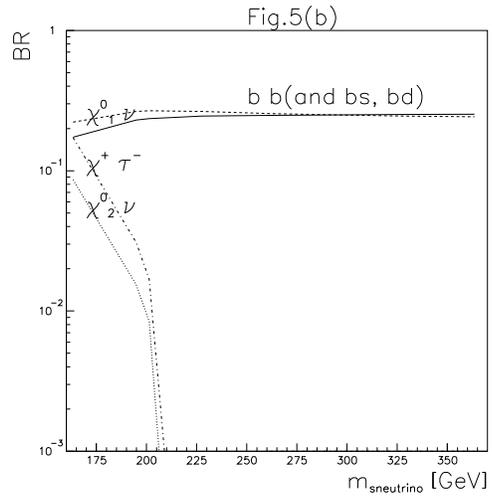}}
\end{center}
\caption{\label{fig5}
Branching ratios of sneutrino $\tilde\nu_3$ with
        (a) $\lambda^{'}_{333}=\lambda^{'}_{332}=
             \lambda^{'}_{323}=0.45$ and with
        (b) $\lambda^{'}_{333}=\lambda^{'}_{332}=
             \lambda^{'}_{323}=0.1$.}
\end{figure}

\begin{figure}[t] 
\leavevmode
\begin{center}
\mbox{\epsfxsize=7.truecm\epsfysize=7.truecm\epsffile{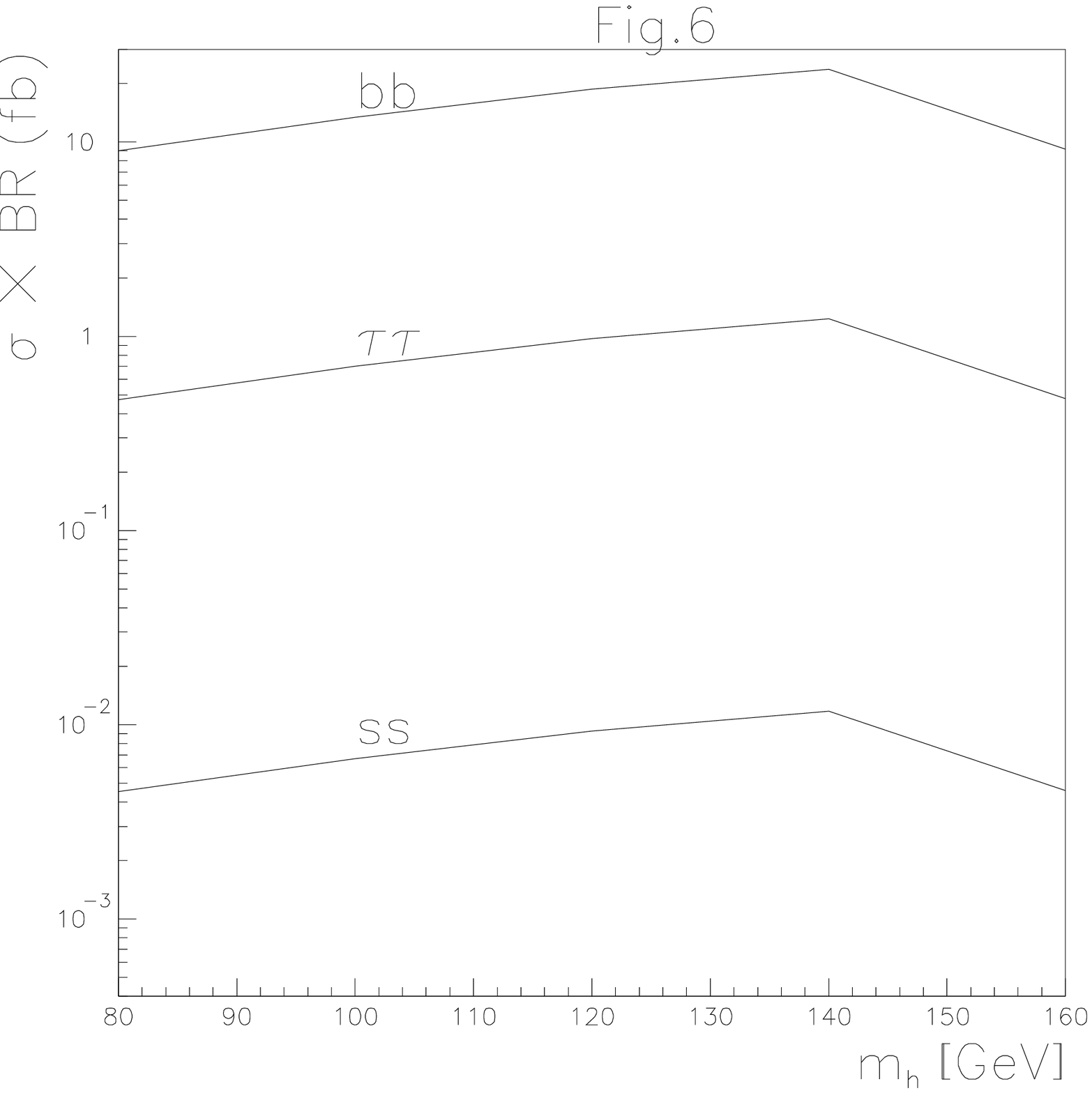}}
\end{center}
\caption{\label{fig8}
As a function of the Higgs mass, $\sigma(\gamma\gamma\rightarrow h^0)
\times BR(f\bar f)$, where $f=b,\,\tau$ or $s$, as denoted in the
figure.}
\end{figure}

\par
However, we need consider other decay modes if the $R_p$-violating
parameters are very small. In Fig. 4, we can see that $\tilde{\nu}\rightarrow
\tilde{\chi}^0_1 +\nu$ will dominate if we take $\lambda_{233}=0.01$.
In this case, we should detect $\tilde{\chi}_1^0$ with its
$R_p$-violating decay. From the $\lambda$ terms, $\tilde{\chi}^0_1
\rightarrow 2 l+\nu$ will dominate,
and in the $\lambda^{'}$ case,
we have $\tilde{\chi}^0_1 \rightarrow 2\ jets+(\nu, lepton)$,
as shown in Fig.7 (a).

\begin{figure}[t]
\leavevmode
\begin{center}
\mbox{\epsfxsize=7.truecm\epsfysize=7.truecm\epsffile{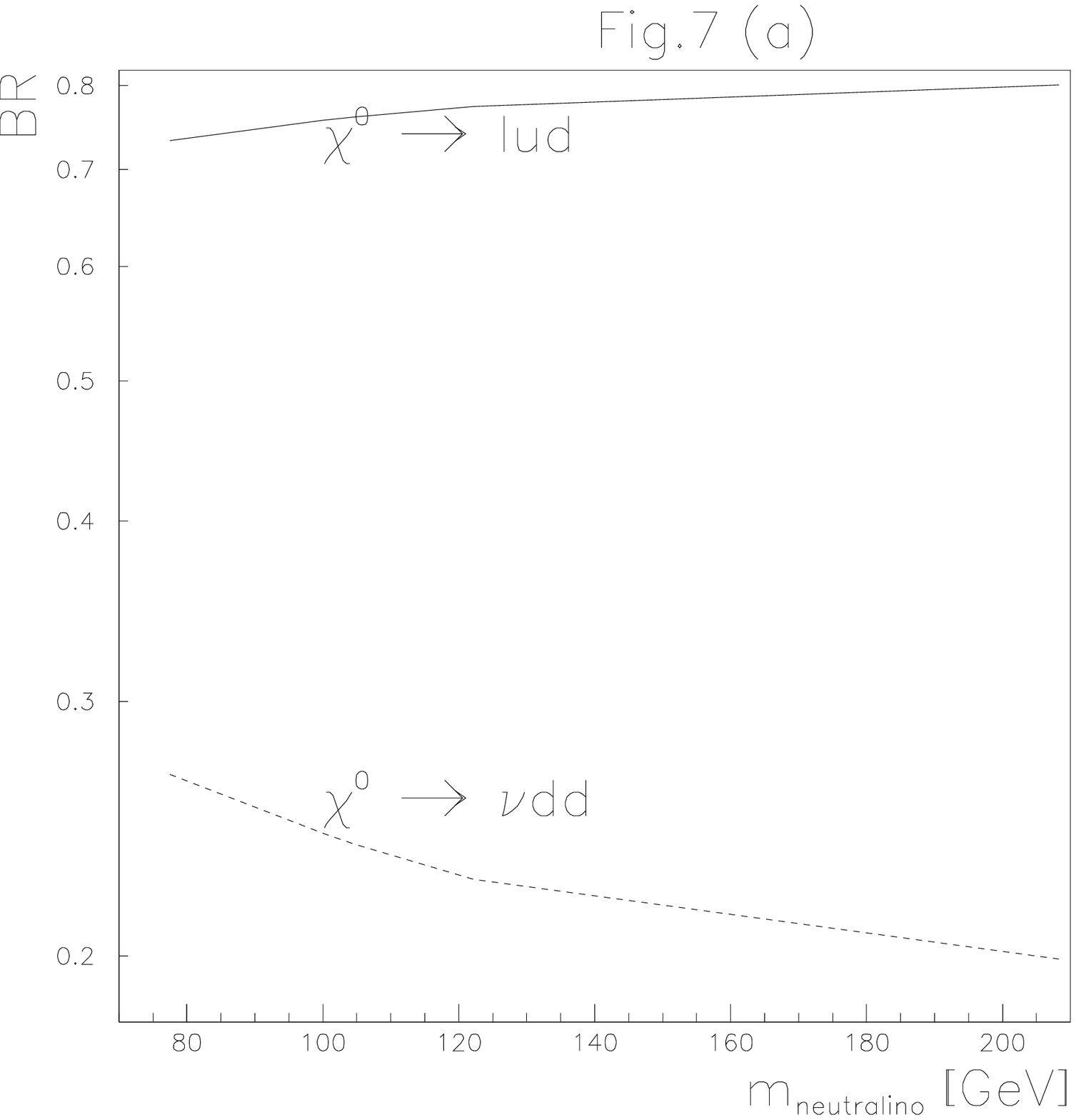}}
\mbox{\epsfxsize=7.truecm\epsfysize=7.truecm\epsffile{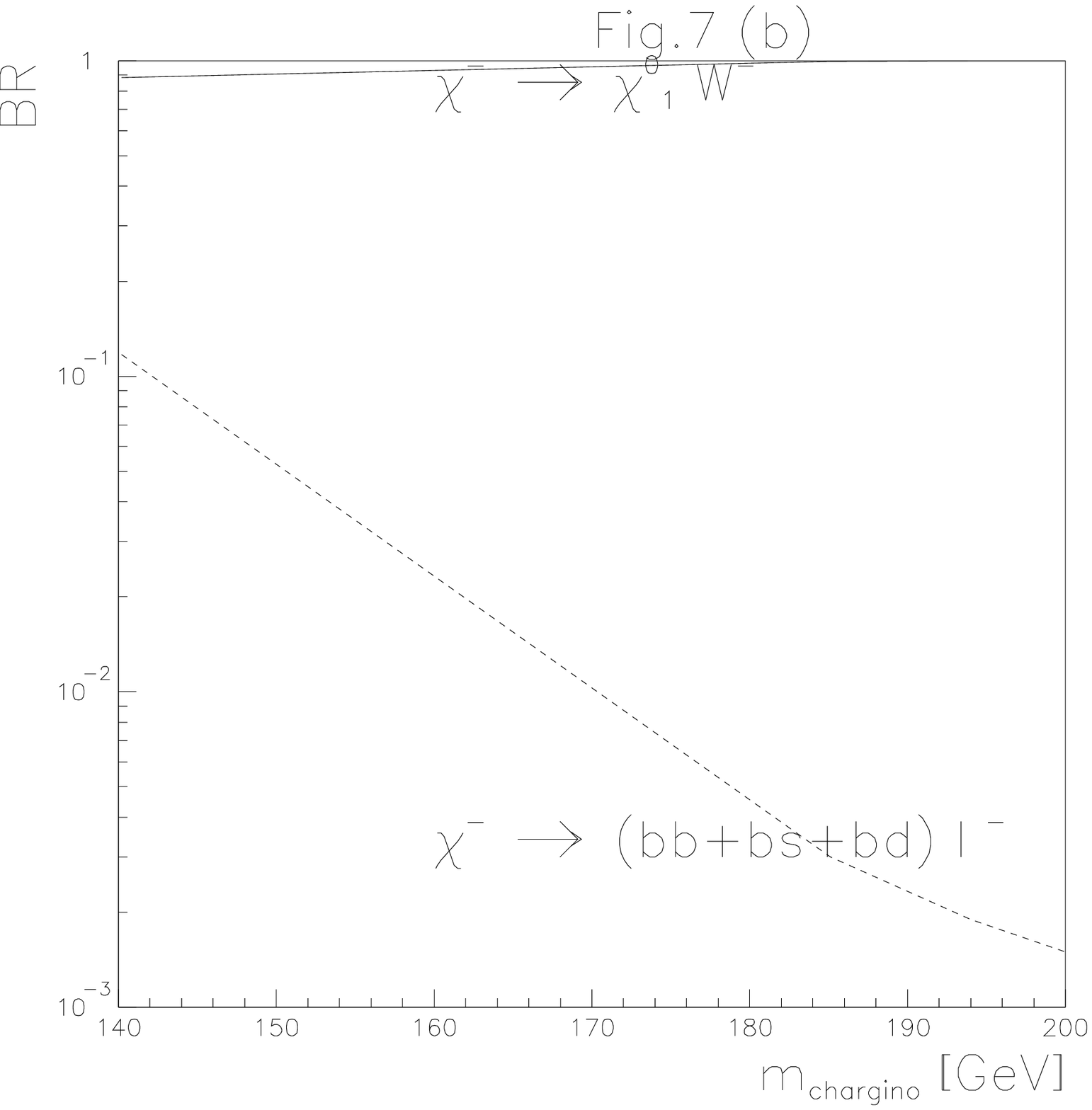}} 
\end{center}
\caption{\label{fig6}
(a) Branching ratio of 
            neutralino $\tilde{\chi}_1^{0}$
            via $\lambda^{'}$ terms.  
(b) Branching ratio of chargino $\tilde{\chi}_1^{\pm}$, 
with
            $\lambda^{'}_{333}=\lambda^{'}_{322}=0.45$.}
\end{figure}

{\it Decay of charginos $\tilde{\chi}_1^{\pm}$.}
The lightest chargino can decay directly to sneutrinos
or sleptons if it is heavier than the corresponding thresholds.
In our case, chargino decay to the lightest neutralino and $W$ boson 
(real or virtual) will dominate with chargino mass below 200 GeV, as 
shown in Fig. 7 (b).
Combined with the neutralino decay (shown in Fig.~7),
the chargino $\tilde{\chi}^{\pm}_1$, which is produced with
other lepton, could be detected at the LC with multi-lepton
signals.

\par
\begin{flushleft} {\bf 4. Conclusion} \end{flushleft}
\par
\noindent
We have studied the single production and decay of sneutrino and the lightest
chargino in $\gamma\gamma$ collisions.
The cross section for the production of sneutrinos
with mass below 400 GeV in the future LC experiments with c.m. energy 500
GeV is above 0.005 fb with $\lambda_{233}=0.1$ and 0.05 fb with
$\lambda^{'}_{333}=0.45$, allowed by experimental limits.
If we cannot find any signals from the experiments, we could improve the
present upper bounds on $\lambda$ and $\lambda^{'}$ or exclude sneutrino
with mass below 400 GeV.

The single production of charginos in photon colliders through
$R_p$ violating couplings is observable only if sneutrino and stau
are light.
The nondiagonal decay channels are important for detection.
The cross section for the processes are at the observable level when
the chargino is lighter than 200 GeV assuming the present $R_p$-violating
limits.

\begin{flushleft} {\bf Acknowledgement} \end{flushleft}

\noindent
Z.-H. Yu thanks the World Laboratory, Lausanne, for the scholarship.

\newpage
\begin{center} {\bf Appendix}\end{center}

\par
\noindent
{\bf A. Loop integrals:}
\par
\noindent
We adopt the definitions of two-, three-, and four-point one-loop
Passarino-Veltman integral functions of reference \cite{s19}\cite{s20}.
The integral functions are defined as follows:
\par
The two-point integrals are:
$$
\{B_0;B_{\mu};B_{\mu\nu}\}(p,m_1,m_2)=
{\frac{(2\pi\mu)^{4-n}}{i\pi^2}}\int d^n q
{\frac{\{1;q_{\mu};q_{\mu}q_{\nu}\}}{[q^2-m_1^2][(q+p)^2-m_2^2]}},
\eqno(A.a.1)
$$
The function $B_{\mu}$ should be proportional to $p_{\mu}$:
$$
B_{\mu}(p,m_{1},m_2)=p_{\mu} B_{1}(p,m_1,m_2)
\eqno(A.a.2)
$$
Similarly we get:
$$
B_{\mu\nu}=p_{\mu}p_{\nu} B_{21}+g_{\mu\nu} B_{22}
\eqno(A.a.3)
$$
We denote $\bar{B}_{0}= B_{0}-\Delta$, $\bar{B}_{1}= B_{1}+\frac{1}{2}\Delta$
and $\bar{B}_{21}= B_{21}-\frac{1}{3}\Delta$. with $\Delta= \frac{2}{\epsilon}
-\gamma +\log (4\pi)$, $\epsilon=4-n$. ${\mu}$ is the scale parameter.
The three-point and four-point integrals can be obtained similarly.
\par
The numerical calculation of the vector and tensor loop integral functions
can be traced back to the four scalar loop integrals $A_0$, $B_0$, $C_0$
and $D_0$ in Ref. \cite{s19}, \cite{s20} and the references therein.
\par
\vskip 0.5 in
\noindent
{\bf B. Sparticle masses}
\par
\noindent
The supersymmetric parameters which we use in our calculations
are shown in the following table as well as the resulting sparticle masses:

\noindent
{\bf Table 1 } We take here $m_{0}=100$ GeV,
                                $A_0=-100$ GeV, $\tan \beta =3$ and
                                $sign(\mu)=+$. The masses are given 
                                in GeV units.
\begin{center}
\begin{tabular}{|c|c|c|c|c|c|c|c|} \hline
$m_{1/2}$ & $m_{\tilde{t}_2}$ & $m_{\tilde{\tau}_2}$ & $m_{\tilde{\nu}}$&
$m_{\tilde{\chi}^0_1}$ &$m_{\tilde{\chi}^{\pm}_1}\sim m_{\tilde{\chi}_2^0}$ \\ \hline
200 & 515 & 179 &163 & 78 & 140  \\ \hline
250 &613&208&195&100&185\\ \hline
260&633&214&201&105&194 \\ \hline
300 &713&239&228&122&229 \\ \hline
350&815&271&261&144&273 \\ \hline
400&917&304&295&166&316 \\ \hline
450&1021&337&329&187&359 \\ \hline
500&1125&371&364&208&402 \\ \hline
\end{tabular}
\end{center}

\end{document}